\documentclass[twocolumn]{aastex6}

\newcommand{\cgs}{${\rm erg}\,{\rm cm}^{-2}\,{\rm s}^{-1}$\,}

\newcommand{\lat}{\textit {Fermi}--LAT}

\slugcomment{}
\shorttitle{M33 and Arp~299}
\shortauthors{Xi et al.}

\begin{document}

\title{GeV $\gamma$-ray emission from M33 and Arp~299}
\author{Shao-Qiang Xi\altaffilmark{1,2}, Hai-Ming Zhang\altaffilmark{1,2}, Ruo-Yu Liu\altaffilmark{1,2}, Xiang-Yu Wang\altaffilmark{1,2}}
\altaffiltext{1}{School of Astronomy and Space Science, Nanjing University, Nanjing 210023, China; xywang@nju.edu.cn}
\altaffiltext{2}{Key laboratory of Modern Astronomy and Astrophysics (Nanjing University), Ministry of Education, Nanjing 210023, China}

\begin{abstract}
Star-forming galaxies are huge reservoirs of cosmic rays (CRs) and these CRs convert a significant fraction of their
energy into $\gamma$-rays by colliding with the interstellar medium (ISM). Several nearby star-forming galaxies have been detected in GeV-TeV $\gamma$-rays. It is also found that the $\gamma$-ray luminosities in
0.1-100 GeV correlate well with indicators of star formation rates of the galaxies, such as the total
infrared (IR) luminosity. In this paper,  we report a systematic search for possible $\gamma$-ray emission from  galaxies in the  IRAS Revised Bright Galaxies Sample, using 11.4 years of $\gamma$-ray data taken by the {\it Fermi} Large Area Telescope (LAT). Two new galaxies, M33 and Arp~299, {   are detected  significantly}. The two galaxies are consistent with the empirical correlation between the $\gamma$-ray luminosity and total infrared luminosity,  suggesting that their $\gamma$-ray emissions should   mainly originate from CRs interacting with ISM.
Nevertheless, there is a tentative evidence that the flux of the $\gamma$-ray emission from Arp~299 is variable. If the variability is true, part of the emission from Arp 299 should originate from the obscured AGN in this interacting galaxy system. In addition,  we find that the
$\gamma$-ray excess from M33 {   is located at} the northeast region of the galaxy, where  a supergiant H II region,  NGC604, resides. This indicates that some bright star-forming regions in spiral galaxies could play a dominant role in the galaxy  in producing $\gamma$-ray emission.

\end{abstract}


\section{Introduction}
It is  believed that Galactic cosmic rays (CRs) are
accelerated by supernova remnant (SNRs) or massive star clusters in our Galaxy. CR protons
interact with the interstellar gas and produce neutral pions
(schematically written as $p+p\rightarrow \pi^0+{\rm other \,
products}$), which in turn decay into $\gamma$-rays
($\pi^0\rightarrow\gamma+\gamma$). {   Seven star-forming galaxies }have been firmly detected in $\gamma$-rays with the Fermi Large Area Telescope (LAT), including the Large Magellanic Cloud \citep[LMC;][]{2010A&A...512A...7A,2016A&A...586A..71A}, the Small Magellanic Cloud \citep{2010A&A...523A..46A}, the Andromeda galaxy M31 \citep{2010A&A...523L...2A}, starburst galaxies M82 and NGC~253 \citep{2010ApJ...709L.152A}, NGC~2146 \citep{2014ApJ...794...26T} and Arp~220 \citep{2016ApJ...821L..20P,2016ApJ...823L..17G}. In addition, a few star-forming galaxies with
obscured active galactic nuclei (AGNs), such as NGC~1068, NGC~4945, NGC 3424 and UGC 11041,  have been detected by \lat\/ \citep{2012ApJ...755..164A,2019ApJ...884...91P}.

Noting the connection between star formations and CRs
in starburst galaxies, some authors have proposed scaling
relationships between star formation rates (SFRs) and $\gamma$-ray luminosities \citep{2002ApJ...575L...5P,2004ApJ...617..966T,2007ApJ...654..219T,2007APh....26..398S,2010MNRAS.403.1569P,2011ApJ...734..107L}. SFR indicators include the total infrared (IR)
luminosity  in $8-1000\,\mu$m  \citep{1998ApJ...498..541K}, and radio
continuum luminosity at 1.4 GHz produced by synchrotron emitting CR electrons \citep{2001ApJ...554..803Y}. With the accumulation
of \lat\/ data, the correlation
between $\gamma$-ray luminosities and SFR indicators are first
found in \cite{2010A&A...523L...2A}. \cite{2012ApJ...755..164A} studied a sample of 69 dwarf, spiral, and luminous and
ultraluminous IR galaxies using 3 years of data collected by
\lat\/. They find further evidence for quasi-linear
scaling relation between the {   $\gamma$-ray luminosity and total}
infrared luminosity.
{   This correlation was later extended to}
higher luminosity galaxies with $\gamma$-ray detection from a luminous IR galaxy, NGC~2146 \citep{2014ApJ...794...26T}, and an ultraluminous IR galaxy, Arp~220\citep{2016ApJ...821L..20P,2016ApJ...823L..17G}.

In this paper,  we report a systematic search for possible $\gamma$-ray emission from  galaxies in the IRAS Revised Bright Galaxies Sample, using 11.4 years of $\gamma$-ray data {\bf collected} by the \lat\/ telescope. {While the result of  detection of GeV emission from M33 and Arp~299 has been briefly mentioned in our previous paper (\citealt{Xi2020}, hereafter Paper I), here we  present the details of the two $\gamma$-ray sources and discuss the nature of their emissions.} The paper is organized as follows.
In \S 2, we present a description of the galaxy sample selection and the \lat\/ data analysis procedure. In \S 3, we present the results of the analysis. In \S 4, we discuss the nature of the $\gamma$-ray emissions from M33 and Arp~299.

\section{Data set and analysis methods}
The scaling relation reported in \cite{2012ApJ...755..164A} implies the $\gamma$-ray detection is likely associated with bright IR galaxies. We selected our sample galaxies from the IRAS Revised Bright Galaxies Sample\footnote{This is a complete flux-limited sample of all extragalactic objects  brighter than 5.24\,Jy at 60\,$\mu m$, covering the entire sky surveyed by IRAS at Galactic latitudes $ |b|> 5^{\circ}$.} \citep{2003AJ....126.1607S}, excluding the 15 IR-bright galaxies that have been detected in $\gamma$-rays with  \lat\/ and listed  in \lat\/ Fourth Source Catalog \citep[4FGL;][]{2020ApJS..247...33A}. We performed the standard sequence of analysis steps for each galaxies (described in our Paper I), resulting in the detection of two new $\gamma$-ray sources that are, respectively, spatially coincident with M33 and Arp~299. The details of the analysis for these two galaxies are given as below.

\lat\/ is a pair-conversion telescope covering the energy range from 20~MeV to more than 300~GeV with a field of view of 2.4~sr \citep{2009ApJ...697.1071A}. For the analysis in this work, we employed recent developments of the Science Tools and use the \lat\/ Pass 8 Source class events collected in $\sim 11.4$ yr, which include {   both front and back-converted LAT events and correspond to P8R3\_SOURCE\_V2 instrument response functions}, but exclude the events with a zenith angle larger than $90^\circ$ in order to remove the contaminant from the Earth limb.

For the galaxy M33, we selected the events in the energy range $0.3-500$\,GeV and within a rectangular {\bf region of interest} (ROI) of size $17^\circ \times 17^\circ$ centered at M33 IR center ($\alpha_{2000}=23.475^\circ,\delta_{2000}=30.669^\circ$). We used $gtmktime$ tool to select time intervals expressed by (DATA\_QUAL $> 0$) \&\& (LAT\_CONFIG ==1), and binned the data in 20 logarithmically spaced bins in energy and in a spatial bin of $0.025^\circ$ per pixel The $\gamma$-ray background model consists of the latest template $\rm gll\_iem\_v7.fits$ for Galactic interstellar emission and the isotropic template  with a spectrum described by the file $\rm iso\_P8R3\_SOURCE\_V2\_v01.txt$, as well as the sources listed in the 4FGL catalog within $20^\circ$ around M33. One possible shortcoming of using the 4FGL catalog (based on 8 years of LAT observations) to perform the search within a data set covering 11.4 yr is that unrelated new point sources may be discovered inside the ROI of the target source, which may influence the analysis. We produce a $6^\circ \times 6^\circ$ map of the Test Statistic (TS)  \footnote{TS is defined as $\rm{TS}=-2({\rm ln}L_0-{\rm ln}L)$, where $L_0$ is the maximum-likelihood value for null hypothesis and $L$ is the maximum-likelihood with the additional point source with a power-law spectrum.} centered at M33 {\bf IR} center
 to search for the new background $\gamma$-ray sources. \footnote{  { For the new $\gamma$-ray point source, the best location and uncertainty can be determined by maximizing the likelihood value with respect to its position only and using the distribution of Localization Test Statistic (LTS), defined by twice the logarithm of the likelihood ratio of any position with respect to the maximum. Using the  $6^\circ \times 6^\circ$ TS map  with the resolution of $0.05^\circ$ per pixel, we determined the position of the new sources at the pixels of the local peak TS value preliminarily.  For each new sources, we produced the $0.6^\circ \times 0.6^\circ$ TS and LTS map with the resolution of $0.01^\circ$ per pixel around the preliminary position to determine the best-fit location and uncertainty, respectively}}. The criteria of the new {\bf background} source is that the  $\gamma$-ray excess has a significance of $\rm TS > 25$ above the diffuse background and has an angular separation larger than $0.3^\circ$ from the center of M33. We find three new background sources and include them in our background model for M33 (see the appendix A).

For the galaxy Arp~299, we selected the events in the energy range $0.3-500$\,GeV within a rectangular ROI of size $17^\circ \times 17^\circ$ centered at the galaxy IR center ($\alpha_{2000}=172.136^\circ,\delta_{2000}=58.561^\circ$).  The rest steps, i.e., the data filter, data bin and background modeling are carried out with similar approaches to what have been done for M33. {There is a new background source  in the  $6^\circ \times 6^\circ$ region centered at Arp~299}.

In the likelihood analysis, we allow each source within $6.5^\circ$ from the ROI center to have a free normalization (the $68\%$ containment radius of photons at normal incidence with an energy of 300~MeV is roughly $2.5^\circ$). This choice ensures that $99.9\%$ of the predicted $\gamma$-ray counts is contained within the chosen radius. The normalizations of the Galactic and isotropic diffuse components are always left free. {   For the background-only fitting of M33 and Arp 299, we first free the spectral parameter (including normalization and index) of sources within $6.5^\circ$ region around the galaxies and then fixed the index for subsequent analysis.}

\section{Data analysis results}

 \begin{table*}
\begin{deluxetable}{lcccccc}
\tabletypesize{\small}
\tablecaption{Spatial template analysis and spectral results of M 33 and Arp 299\label{tab:1}}
\tablewidth{0pt}
\tablehead{
\colhead{Spatial model} &\colhead{$\rm TS$}  &\colhead{$R.A.$}&  \colhead{(Decl.)} & \colhead{$F_{\rm 0.1-100 \rm GeV}$}& \colhead{$\Gamma$} &\colhead{$N_{\rm dof}$}  \\ 
                                       &                          &    \colhead{[deg]}&    \colhead{[deg]} & \colhead{[$10^{-12}$\,\cgs]}         &                                      &  \\
              \colhead{(1)} &   \colhead{(2)} &  \colhead{(3)} &  \colhead{(4)} &  \colhead{(5)} &  \colhead{(6)} &  \colhead{(7)}}
\startdata
\multicolumn{7}{c}{M33}\\
\hline%
Point source (free)\tablenotemark{a} & 25.1 & 23.609  & 30.784 & 1.28$\pm$0.42 & 2.23$\pm$0.24 & 4 \\
Point source (fixed)\tablenotemark{b}          & 16.7   & 23.475 & 30.669 &   1.34$\pm$0.47&2.41$\pm$0.26 & 2 \\
$0.23^\circ$ Disk \tablenotemark{c}           & 23.2   & 23.475 & 30.669 &   1.55$\pm$0.35&2.22$\pm$0.42 & 3\\
Herschel/PACS map (160 $\mu {\rm m}$))          & 22.8   & & &   1.48$\pm$0.40&2.22$\pm$0.42 & 2 \\
IRAS map (60 $\mu {\rm m}$))                            & 23.9   & & &   1.52$\pm$0.40&2.20$\pm$0.43 & 2 \\
\hline%
\multicolumn{7}{c}{Arp~299}\\
\hline%
Point source (free)\tablenotemark{a} & 27.8 & 172.050  & 58.526 & 1.08$\pm$0.28 & 2.07$\pm$0.20 & 4 \\
\enddata
\tablenotetext{a}{   The point source is at the best-fit location. The best-fit location is determined at the position of the peak TS value using the TS map with the resolution of $0.01^\circ$ per pixel.}
\tablenotetext{b}{   The point source is  at the {\bf IR} center of the {\bf galaxy M33}.}
\tablenotetext{c} {   The center of the disk model are fixed to the {IR} center of the {\bf galaxy M33}.}
\tablecomments{(1)Spatial model name;
(2)$\rm TS$ value for each spatial model;
(3) right ascension J2000;
(4) declination J2000;
(5) 100 MeV-100 GeV $\gamma-ray$ average flux;
(6) power-law spectral photon index derived by broad band spectrum fitting;
(7) degree of  freedom for each spatial model .}
\end{deluxetable}
\end{table*}

\subsection{M33}
\subsubsection{Morphological analysis}
\begin{figure}[htbp]
\includegraphics[width=0.45\textwidth]{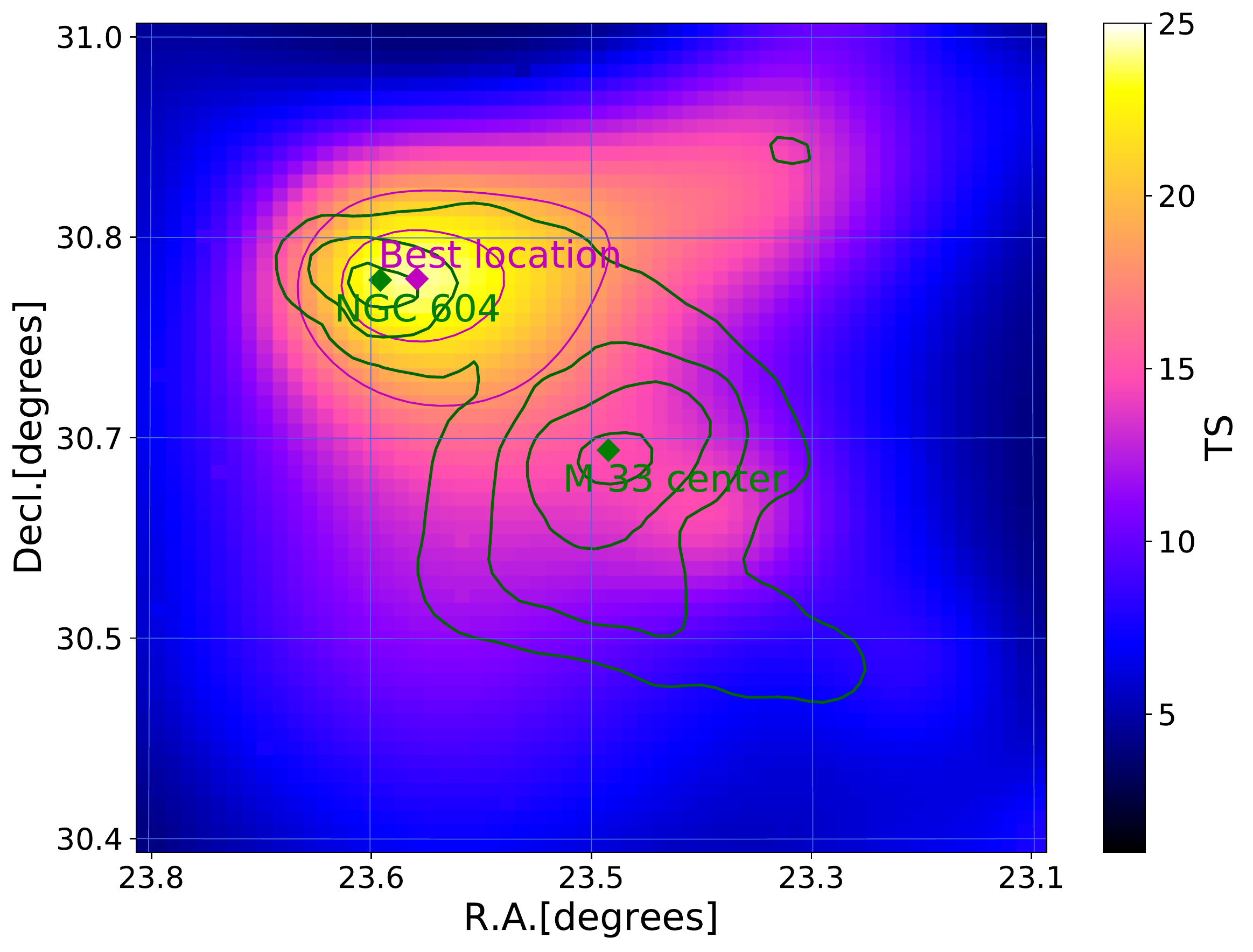}
\caption{TS map in the energy band $0.3-500$\,GeV around M33. The purple contours correspond to $68\%$ and $95\%$ confidence region assuming a template of a point-like source at the best-fit location. The dark green contours correspond to the map of IR flux  measured by IRAS at $60\mu$m. }
\label{fig:1}
\end{figure}

Fig.\ref{fig:1} shows the $0.6^\circ\times 0.6^\circ$ TS map  in $0.3-500$\,GeV  around M33. We find that the position of the TS peak locates at the northeast part of the galaxy. {   We tested various spatial models and investigated which model is better by employing $\Delta \rm TS = TS_{1}-TS_{2}$, where $\rm TS_{1}$ and $\rm TS_{2}$ are respectively the TS value of  model 1 and  model 2 respectively \footnote{{Considering the difference in degree of freedom of the model 1 and model 2 is $\Delta N_{dof}$, we can expect the cumulative density of  $\Delta \rm TS$ follows a $\chi^2$ distribution with $\Delta N_{dof}$ degree of freedom. In the case of the same number of degrees of freedom, we simply assume the $\chi^2$ distribution with one degree of freedom.}}}. We first explored the point source models at this best-fit location (i.e., the position of the peak TS value) and  at the center of M33, respectively. The TS values are, respectively, 25.1 and 16.7, which suggests that the source is likely to be offset from the galaxy center with a significance $2.4\sigma$. In addition, we considered spatially extended templates based on Herschel/PACS map at $160\,\mu$m and IRAS map at $60\, \mu$m. These templates are used to test the spatial correlation of the $\gamma$-ray emission with star formation sites. The Hershcel/PACS and IRAS map models provide better fits ($\sim 2.5\sigma$) to the data than the point source model at the center of M33, but give almost equally good  fits as the point source model at the best-fit location. We also test the uniform-brightness disk model with free radius centered at the optical center of M33. The TS value peaks when the radius is  $0.23^\circ$. We do not find any improvement over the point source model at the best-fit location.  The results for all the considered morphological tests are shown in Table \ref{tab:1}.

\subsubsection{Flux variability}\label{M33_FV}
\begin{figure}[htbp]
\includegraphics[width=0.45\textwidth]{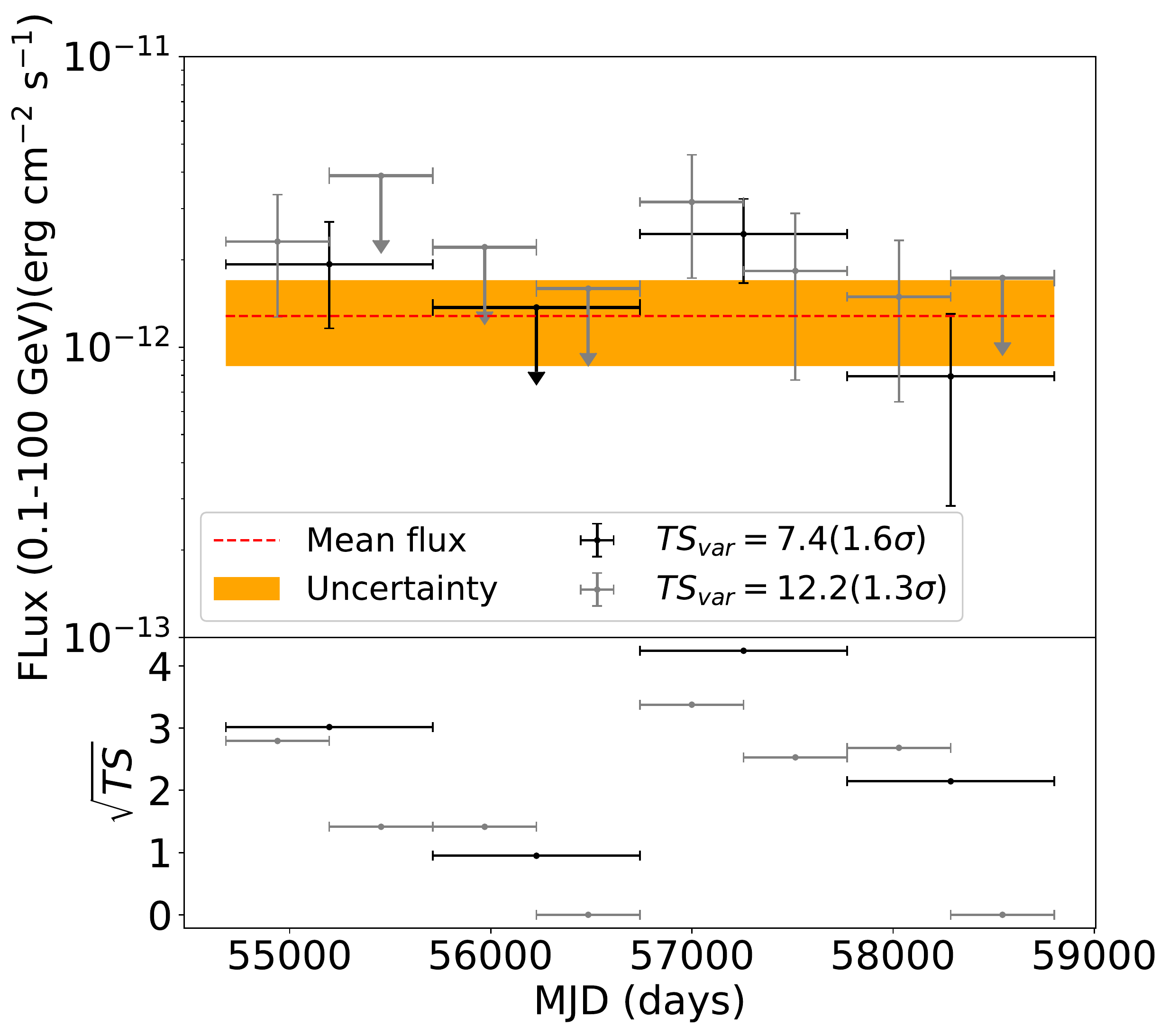}
\caption{Light curves of M33 with 4 and 8 time bins. The mean flux is the averaged flux over the $\sim11.4$ year analysis. The upper limits at $95\%$ confidence level are derived when the TS value for the data points is lower than 4.}
\label{fig:2}
\end{figure}

We retained the point source model at the best-fit location for examining the variability of the $\gamma$-ray flux. We computed  light curves in four and eight time bins over 11.4 yr, for events in the energy range $0.3-500$\,GeV. For the analysis in each time bin, all sources within $6.5^\circ$ region around M33 have their spectra fixed to the shapes obtained from the above broad band analysis. The result is shown in Figure \ref{fig:2}. We then used a likelihood-based statistic to test the significance of the variability. Following the definition in 2FGL \citep{2012ApJS..199...31N} , the variability index from the likelihood analysis is constructed, with a value in the null hypothesis where the source flux is constant across the full time period, and the value under the alternate hypothesis where the flux in each bin is optimized: $TS_{var} = \sum_{i=1}^{N} 2 \times (Log(L_i(F_i))-Log(L_i(F_{mean})))$, where $L_i$ is the likelihood corresponding to bin $i$, $F_i$ is the best-fit flux for bin $i$, and $F_{mean}$ is the best-fit flux for the full time assuming a constant flux. We get $1.3-1.6 \sigma$ significance for the flux variability for the analyses using the above two time bins, which suggests no significant variability for the $\gamma$-ray emission from M33.

\subsubsection{Spectral Analysis}\label{M33_SA}
\begin{figure}[htbp]
\includegraphics[width=0.45\textwidth]{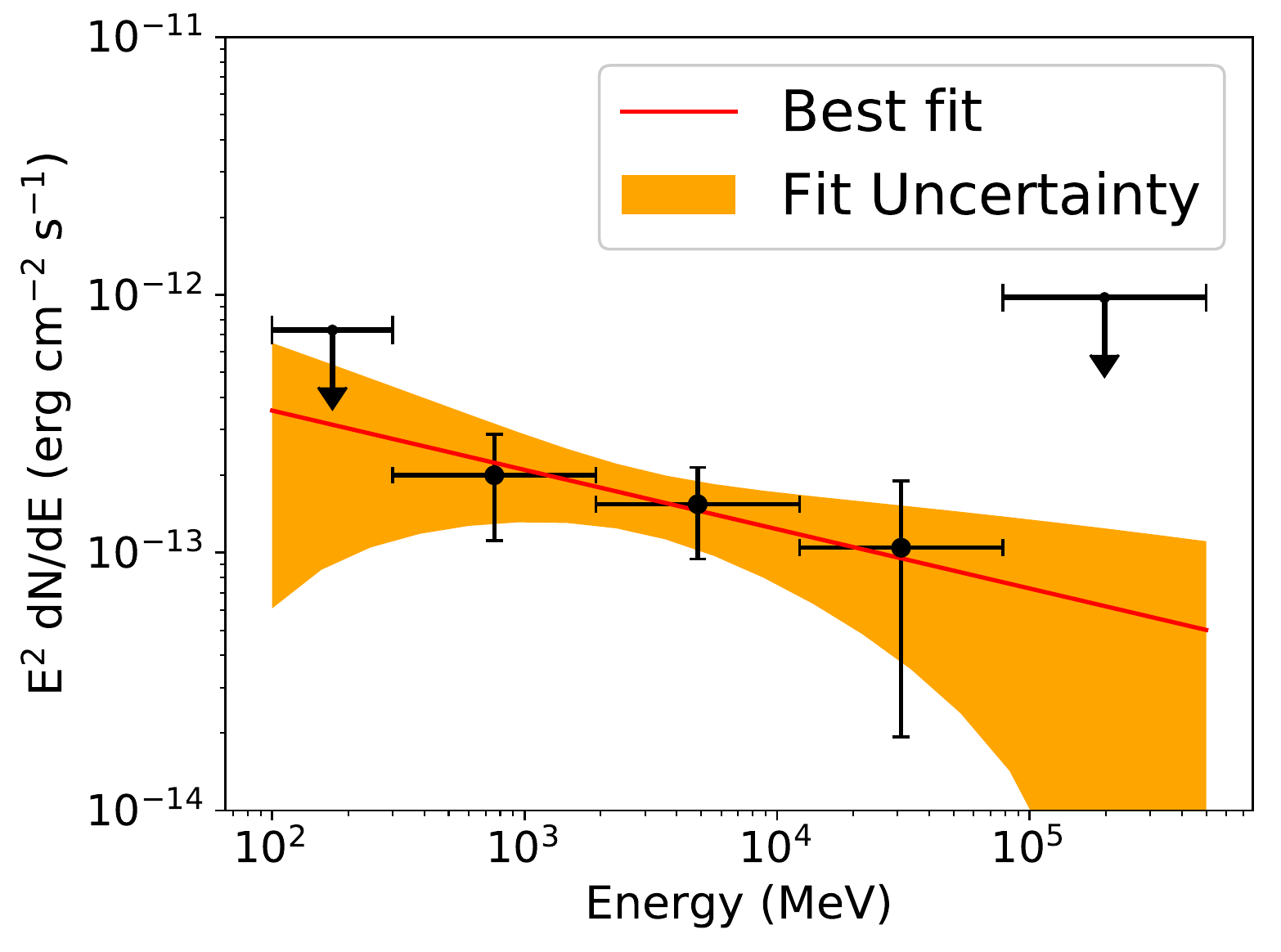}
\caption{SED for M33 in the energy band 0.1-500 GeV. The red line and yellow band represent the wide-band best-fit power-law spectrum and the uncertainty {\bf obtained in the energy band 0.3-500 GeV.}  The upper limits at $95\%$ confidence level are derived when the TS value for the data point is lower than 4 ($<2\sigma$).  }
\label{fig:3}
\end{figure}

For the spectral analysis of M33, we performed a binned maximum likelihood fitting in the $0.3-500$\,GeV energy range with 20 logarithmic energy bins in total. The power law indices are consistent with each other for the four spatial models, as shown in Table \ref{tab:1}. {   We  generated the spectral points based on a maximum likelihood analysis with 4 logarithmic energy bins over $0.3-500$\,GeV.} Within each bin, we used the point source model at the best-fit location and the power law spectrum with a fixed photon index of ${\rm \Gamma = 2}$ and a free normalization. For the background diffuse components and sources within $6.5^\circ$ of M33, we fixed their spectral indices to the best-fit values obtained from the above background fitting, but allowing the normalization to vary. {   We also obtain the spectral point at the energy band  $0.1-0.3$\,GeV to check whether the power-law model is good. We find that the best-fit wide-band power-law model  is consistent with all the spectral points, as shown in Fig \ref{fig:3}}.

\subsection{Arp~299}
\subsubsection{Morphological analysis}
\begin{figure}[htbp]
\includegraphics[width=0.45\textwidth]{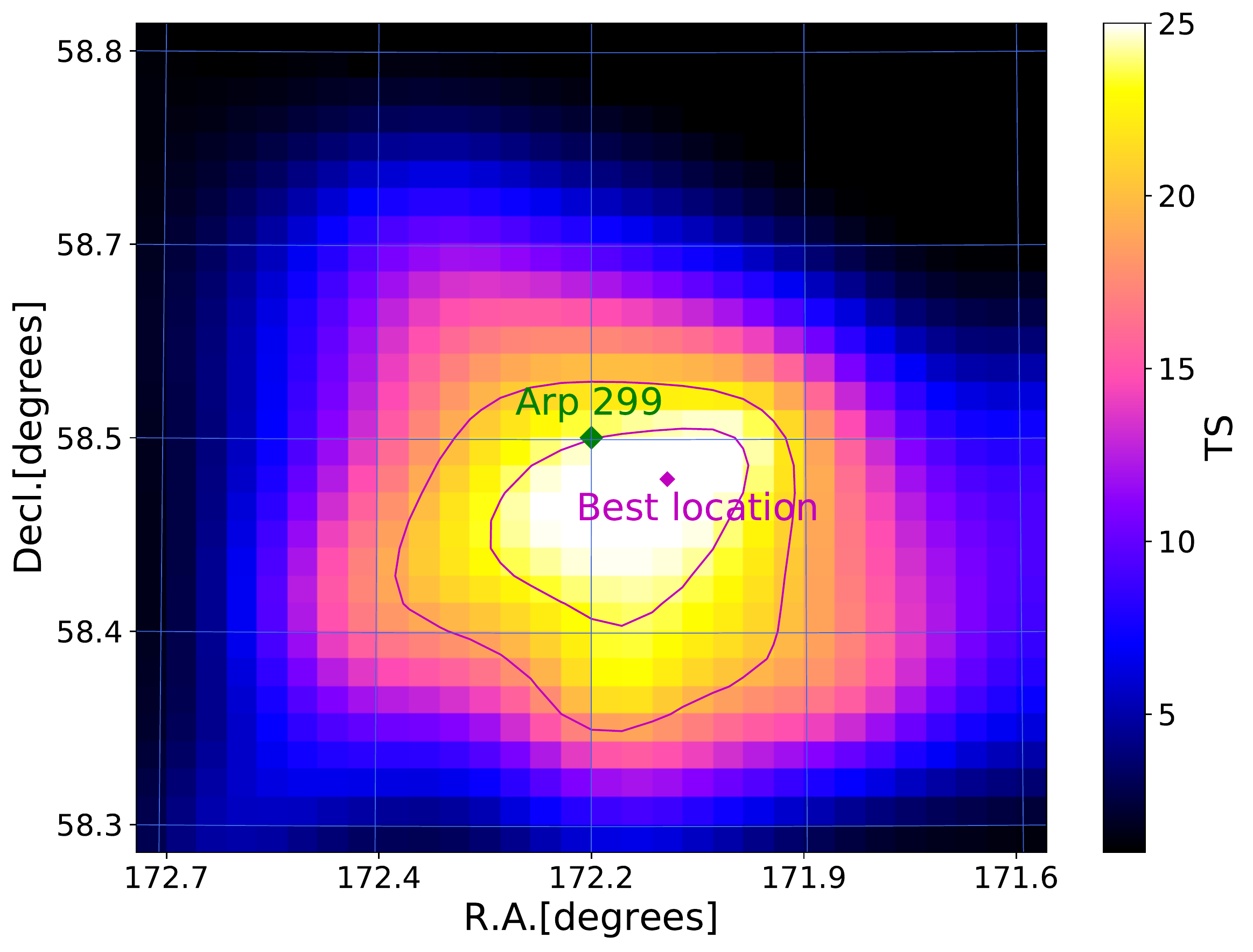}
\caption{TS map in the energy band $0.3-500$\,GeV for Arp~299. The purple contours correspond to $68\%$ and $95\%$ confidence region assuming the point-like source template. {   The dark green diamond  correspond to the optical localization of Arp~299.)}}
\label{fig:4}
\end{figure}
{Fig. \ref{fig:4} shows the TS map in the $0.3-500$\,GeV energy range around the galaxy  Arp~299.}
We find that the galaxy position is located within the $68\%$ confidence region of the $\gamma$-ray excess.

\subsubsection{Flux variability}
\begin{figure}[htbp]
\includegraphics[width=0.45\textwidth]{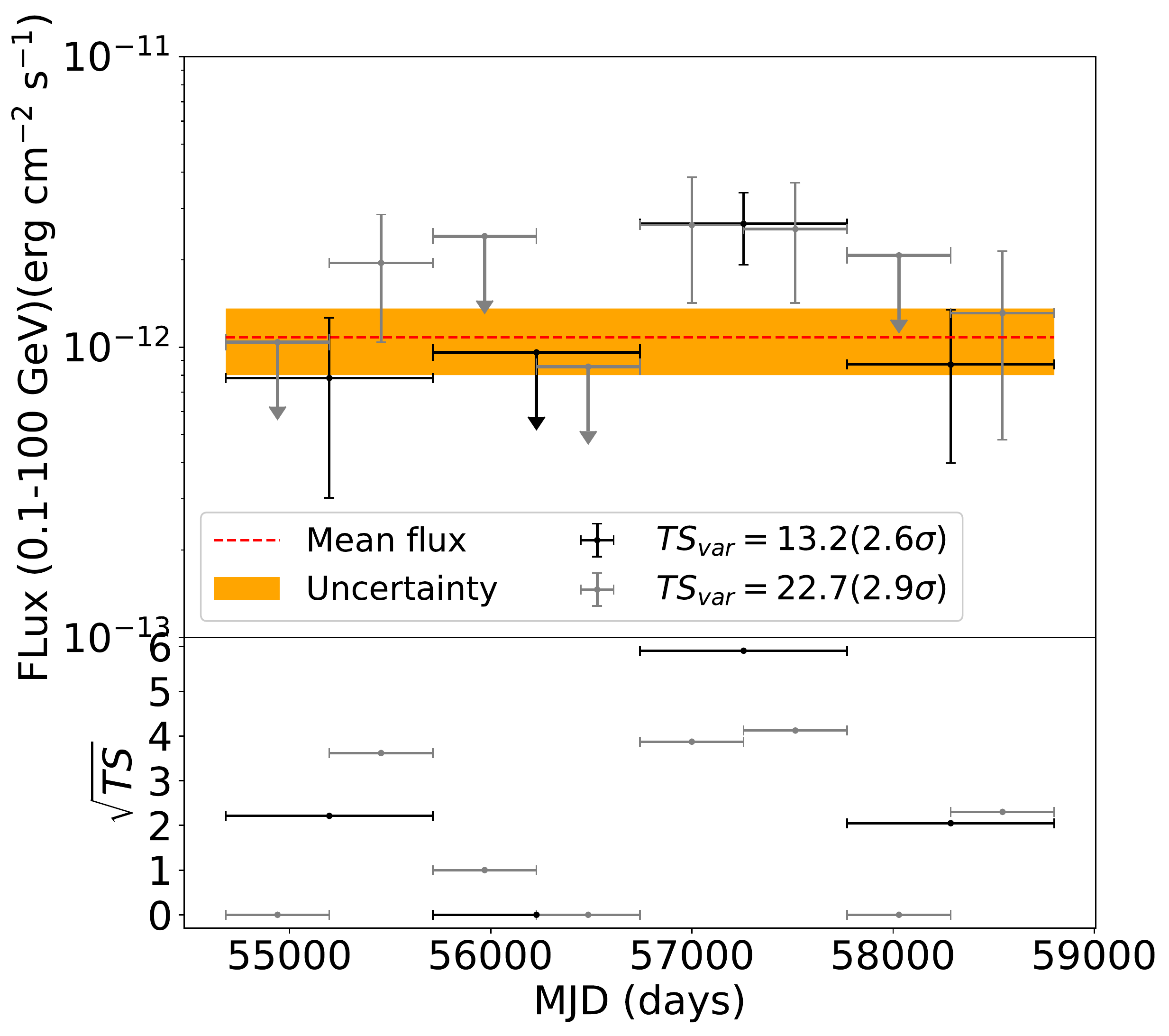}
\caption{ Light curves of Arp~299 with 4 and 8 time bins. The mean flux is the averaged flux over the $\sim11.4$ year analysis. The upper limits at $95\%$ confidence level are derived when the TS value for the data points is lower than 4.}
\label{fig:5}
\end{figure}

To examine the variability of the $\gamma$-ray flux from Arp~299, we computed light curves in four and eight time bins over 11.4 yr for events in the energy range $0.3-500\,$GeV. We followed a procedure similar to that used for M33 in Section \ref{M33_FV}, and the result is shown in Figure \ref{fig:5}.  We obtain a variability significance of $2.6-2.9 \sigma$ for the analyses using the above two time bins, which suggests a mildly significant variability in the $\gamma$-ray emission of Arp~299. We also checked the flux variability with 16  and 40 time bins. We get $3.0 \sigma$ significance for the flux variability using the 16 time bins, which agrees with the analysis using four and eight time bins. For the 40 time-bin case, we get $0.7 \sigma$ significance for the flux variability, which seems to indicate less variability on such short timescales. However, we note that,  for such a weak $\gamma$-ray source,   the statistics in each bin in the 40 time-bin analysis may be too low for a reliable analysis.
\subsubsection{Spectral Analysis}
\begin{figure}[htbp]
\includegraphics[width=0.45\textwidth]{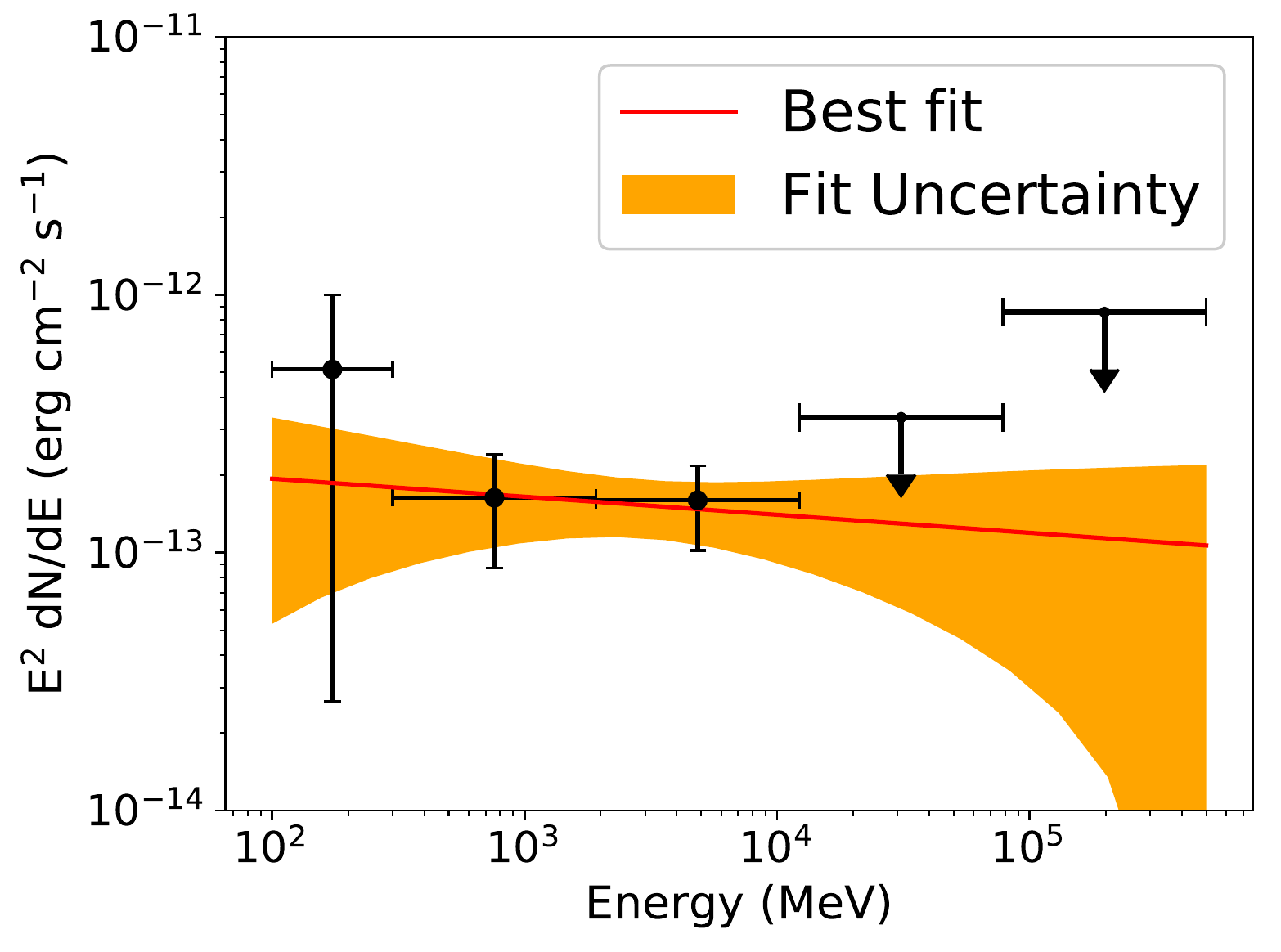}
\caption{{   SED for Arp 299 in the energy band 0.1-500 GeV. The red line and yellow band represent the wide-band best-fit power-law spectrum and the uncertainty {\bf obtained in the energy band 0.3-500 GeV.} }  The upper limits at $95\%$ confidence level are derived when the TS value for the data point is lower than 4. }
\label{fig:6}
\end{figure}

For the spectral analysis of Arp~299, we performed a binned maximum likelihood fitting in the $0.3-500\,$
GeV energy range with 20 logarithmic energy bins in total {   considering the point source model}. The result is shown  in Table.~\ref{tab:1}.  We also generated the spectral points determined by performing a maximum likelihood analysis {\bf in five energy bins} over $0.1-500$\,GeV similar to the case of M33 in Section \ref{M33_SA}. As shown in Figure \ref{fig:6},  the wide-band power-law model is
consistent with these spectral points.

\subsection{Non-detected IR galaxies}
\begin{figure}[htbp]
\includegraphics[width=0.45\textwidth]{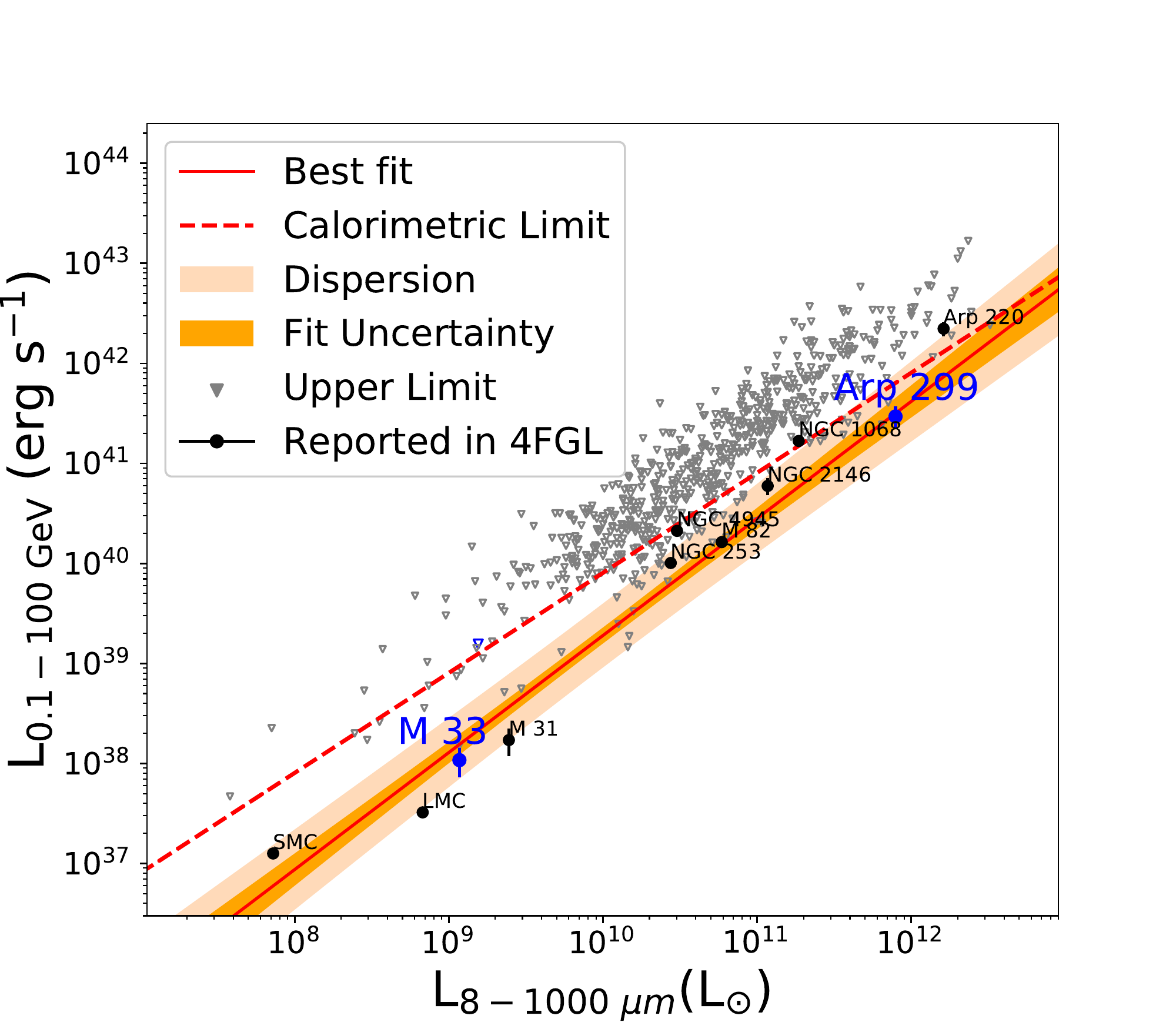}%
\caption{$\gamma$-ray luminosities ($0.1-100$\,GeV) v.s. total IR luminosities ($8-1000 \mu$m) for nearby star-forming galaxies. {The yellow band represents the empirical correlation  ${\rm log}(L_{0.1-100\,{\rm GeV}}/ {\rm erg\,s^{-1}})=(1.17\pm0.07){\rm log}(L_{8-1000\,\mu \rm m}/{10^{10}L_{\odot}})+(39.28\pm0.08)$ with an intrinsic dispersion normally distributed in the logarithmic space and a standard deviation of $\sigma_{D}=0.24$ \citep{2012ApJ...755..164A}. } The $\gamma$-ray luminosities for the galaxies reported in 4FGL are derived from the catalog values \citep{2020ApJS..247...33A}. The IR luminosities are from \cite{2003AJ....126.1607S}. The blue upper downward triangle represents the upper limit for NGC~2403 for the last 5.7-year \lat\/ data.}
\label{fig:7}
\end{figure}

We derived the 95\% C.L. upper limits (UL) for each non-detected galaxy (i.e., ${\rm TS<25}$) using the Bayesian method assuming a power-law spectrum with a fixed photon index of $2.2$. For NGC~2403, we attribute the $\gamma$-ray emission, {   which is present only in the first 5.7-year \lat\/ observation (4 Aug. 2014-25 Mar. 2014), to SN~2004dj (see Paper I). Using the last 5.7-year \lat\/ data (25 Mar. 2014  - 14 Nov. 2019), }we derived an upper limit for NGC~2403 assuming a  point source model at the galaxy center. We compare the ULs on the $\gamma$-ray luminosities ($0.1-100$\,GeV) to the total IR luminosities ($8-1000\,\mu $m) for these non-detected IR galaxies, which is shown in Fig \ref{fig:7}. We find these non-detected IR galaxies are basically consistent with the empirical $L_\gamma-L_{\rm IR}$ correlation.

\section{Discussions and Conclusions}\label{discussion}

\subsection{M33}
As the third largest galaxy in our Local Group, M33 has been considered to be a
promising $\gamma$-ray source due to its proximity and relatively high gas masses and star formation activity. By using nearly 2 years of \lat\/ data, \citet{2010A&A...523L...2A} searched for the $\gamma$-ray emission from M33, but no significant $\gamma$-ray emission was detected. \citet{2017ApJ...836..208A} revisited the $\gamma$-ray emission in the direction M33 using more than 7\,yr of LAT Pass 8 data in the energy range $0.1-100$\,GeV, but still found no significant detection. More recently, \citet{2019ApJ...880...95K}  found positive residual towards the M33 region.  Di Mauro et al. (2019) reported M33  as a  point-like source  with  $TS=39.4$  in the energy band $0.3-1000\rm\ GeV$. However, we find that three new background sources (see the Appendix for details) nearby M33 identified by us are not in their background model.  We repeated our analysis using  the background model excluding these three new background sources and find a similar values of  $\rm TS= 35.9$ and $F_{0.1-100\rm GeV}\sim2.1\times10^{-12}$\,\cgs  for the best-fit point source model. Thus, we think that the difference between our results and that of Di Mauro et al. (2019) is due to the new background sources. 




Our measurement gives a flux of $(1.28\pm0.42) \times 10^{-12}{\rm \,erg\ cm^{-2}\ s^{-1}}$ in the energy range $0.1-100$\,GeV, implying a luminosity of $\sim1.1\times 10^{38} {\rm erg\ s^{-1}}$. In Figure \ref{fig:7}, we show the position of M33 on the  empirical $L_\gamma-L_{\rm IR}$ correlation for local group galaxies and nearby star forming galaxies \citep{2012ApJ...755..164A,2016ApJ...821L..20P}.  M33 agrees well with this correlation, indicating that the $\gamma$-ray emission of M33 may arise from the CR-ISM interaction process.

The TS map of M33  shows that the $\gamma$-ray emission locates at the northeast  region of the galaxy, where a supergiant H~II region, NGC~604, resides. NGC~604 is the second most massive H~II region in the Local Group and it has a relatively high star-formation rate. From the H~I density distribution map of M33,  an over-density of H~I gas filament is seen around that region, with a column density of ${\rm \sim 30 M_\odot \,pc^{-2}}$ \citep{2003ApJS..149..343E}. Such a high-density region provides a thick target for CR-ISM interaction, so the efficiency for the  $pp$ collision is expected to be high.

The energy-loss time of protons due to the $pp$ collision $t_{\text{loss}}$ can be expressed as $\left(
0.5n\sigma _{pp}c\right) ^{-1}$, where 0.5 is the inelasticity, $n$
is the hydrogen atom number density and $\sigma_{pp}$ is the inelastic $pp$
collision cross section. Converting the atom number density to gas
surface density, $\Sigma_{g}=m_{p}nH$, where $m_{p}$ is the mass of proton
and $H$ is the size of the over-density region, the energy-loss time is
\begin{equation}
t_{\text{loss}}=6.5\times 10^{6}\left(\frac{H}{200\text{pc}}\right)\left( \frac{\sigma
_{pp}}{30\text{mb}}\right) ^{-1}\left( \frac{\Sigma _{g}}{30 M_\odot {pc}^{-2}}
\right) ^{-1}\text{yr}\end{equation}
where $H$ is the the typical width of an H~I filament.

CRs are scattered off small-scale
magnetic field inhomogeneities randomly and diffuse out of the H~I filament.
The diffusive escape time is $t_{\text{diff}}=H^{2}/4D$. Here $D=D_{0}\left(
E/E_{0}\right) ^{\delta }$ is the diffusion coefficient, where $D_{0}$ and $%
E_{0}={\rm 3 GeV}$ are normalization factors, and $\delta =0-1$
depending on the spectrum of interstellar magnetic turbulence. The
diffusion time is
\begin{equation}
t_{\text{diff}}=3\times 10^{5}\left( \frac{H}{200\text{pc}}\right)
^{2}\left( \frac{D_{0}}{10^{28}\text{cm}^{2}\text{s}^{-1}}\right)
^{-1}\left( \frac{E_{p}}{3\text{GeV}}\right) ^{-\delta }\text{yr}.
\end{equation}%
With $D_0\sim {10^{27}\text{cm}^{2}\text{s}^{-1}}$,
the escape time is comparable to the $pp$ cooling time, and the region may be considered to be a proton calorimeter. {Although this value is one order of magnitude smaller than the standard diffusion coefficient in the ISM of our Galaxy, it is not {\it a priori} impossible. A recent polarisation analysis on Cygnus-X, a massive star-forming region in our Galaxy, has revealed that the turbulence in the region is dominated by the magnetosonic mode \citep{2018arXiv180801913Z}, which is more effective than the commonly considered Alfv{\'e}nic mode in CR confinement \citep{2002PhRvL..89B1102Y}. There are also a few giant molecular clouds with mass up to $10^6M_\odot$ spatially associated with NGC~604 \citep{2003ApJS..149..343E}, which would enhance the average atom density and subsequently the $\gamma$-ray emissivity by a factor of at least a few in that region. In addition, massive stellar winds are probably efficient CR factories \citep{1980ApJ...237..236C, 1983SSRv...36..173C, 2019NatAs...3..561A}, so we expect the CR density around NGC~604 to be higher than the average CR density in the ISM. This may explain why the peak of $\gamma$-ray emission locates in the northeast region of M33.}

\subsection{Arp~299}
Arp~299 is one of the most powerful merging galaxy system in the local Universe, at a distance of 44\,Mpc \citep{1999ApJ...517..130H}. {The system consists of two galaxies in an advanced merging state, NGC~3690 to the west and IC 694 to the east, plus a small compact galaxy \citep{1999AJ....118..162H}. The total IR luminosity of both galaxies(NGC~3690+IC~694) are ${L_{\rm IR}= 5.16\times 10^{11}L_{\odot}}$ \citep{2002ApJ...571..282C}, so it belongs to the class of Luminous IR Galaxies (LIRGs).} BeppoSAX revealed for the first time the existence of a deeply buried (${N_H=2.5\times10^{24} \rm cm^{−2}}$) AGN with a unabsorbed luminosity of $L_{0.5-100\,\rm keV}= 1.9\times10^{43}{\rm erg~s^{-1}}$ \citep{2002ApJ...581L...9D}. Chandra and XMM-Newton observations later confirmed the existence of a strongly absorbed AGN and located it in the nucleus of NGC~3690, while there is evidence that the second nucleus IC 694 might also host an AGN of lower luminosity \citep{2003ApJ...594L..31Z,2004ApJ...600..634B,2009ApJ...695L.103I,2010A&A...519L...5P,2002ApJ...581L...9D,2013ApJ...779L..14A}. {According to the correlation between the X-ray luminosity ($L_{2-10\,{\rm keV}}$) and the bolometric luminosity ($L_{\rm bol}$) of X-ray selected AGN \citep{2012A&A...545A..45R}, we find an intrinsic luminosity of $L_{\rm bol}\simeq 5\times 10^{43}\rm erg~s^{-1}$ for the obscured AGN. Even if all the AGN luminosity is reprocessed into the IR band, its contribution is negligible to the measured IR luminosity from the galaxy and hence the latter is related to the star-forming process in Arp~299, as in other star-forming galaxies.}

As shown in Fig.\ref{fig:5}, there is a tentative evidence of flux variability in Arp~299. If this variability is true, it may be due to the contribution from the obscured AGN. Some other merging galaxy systems, such as NGC~3424, also show flux variability in $\gamma$-ray emission \citep{2019ApJ...884...91P}. However, different from NGC~3424, Arp~299 lies on the empirical $L_\gamma-L_{\rm IR}$ scaling (see Fig.\ref{fig:7}). {   Considering also that the hint of variability is not very significant($<3\sigma$), this may indicate that the  obscured AGN contributes a subdominant part to the whole $\gamma$-ray flux}.

\subsection{Conclusions}

To summarize, our analysis using 11.4 years of \lat\ data in new detections of $\gamma$-ray emission from M33 and Arp~299. The fluxes of both sources are consistent with the correlation between the $\gamma$-ray luminosities and the total IR luminosities for star-forming galaxies, suggesting that  $\gamma$-ray emissions from the two sources should arise mainly  from CRs interacting with the ISM. However, it is found that there is a tentative evidence of variability in the $\gamma$-ray flux of Arp~299. The variability can be tested in future with longer observation time. If the variability is true, part of the $\gamma$-ray emission should come from the obscured AGN in Arp~299. The morphological analysis of the $\gamma$-ray emission from M33 shows that the peak of the TS map is not located at the galaxy center, but coincident with the supergiant H~II region NGC~604. This implies that
some bright star-forming regions could dominate over the bulk of the galaxy disk in producing  $\gamma$-ray emission.

{\em A note added:} We note that during the final stage of the present work, an independent research paper \citep{Ajello2020} appears online, which conducted a similar study to that of the present work.

\acknowledgments  The work is supported by NSFC grants 11625312 and 11851304, and the National Key
R \& D program of China under the grant 2018YFA0404203.

\clearpage

\appendix
\section{Background model  for M~33 and Arp~299}
\begin{figure}[htbp]
\includegraphics[width=0.45\textwidth]{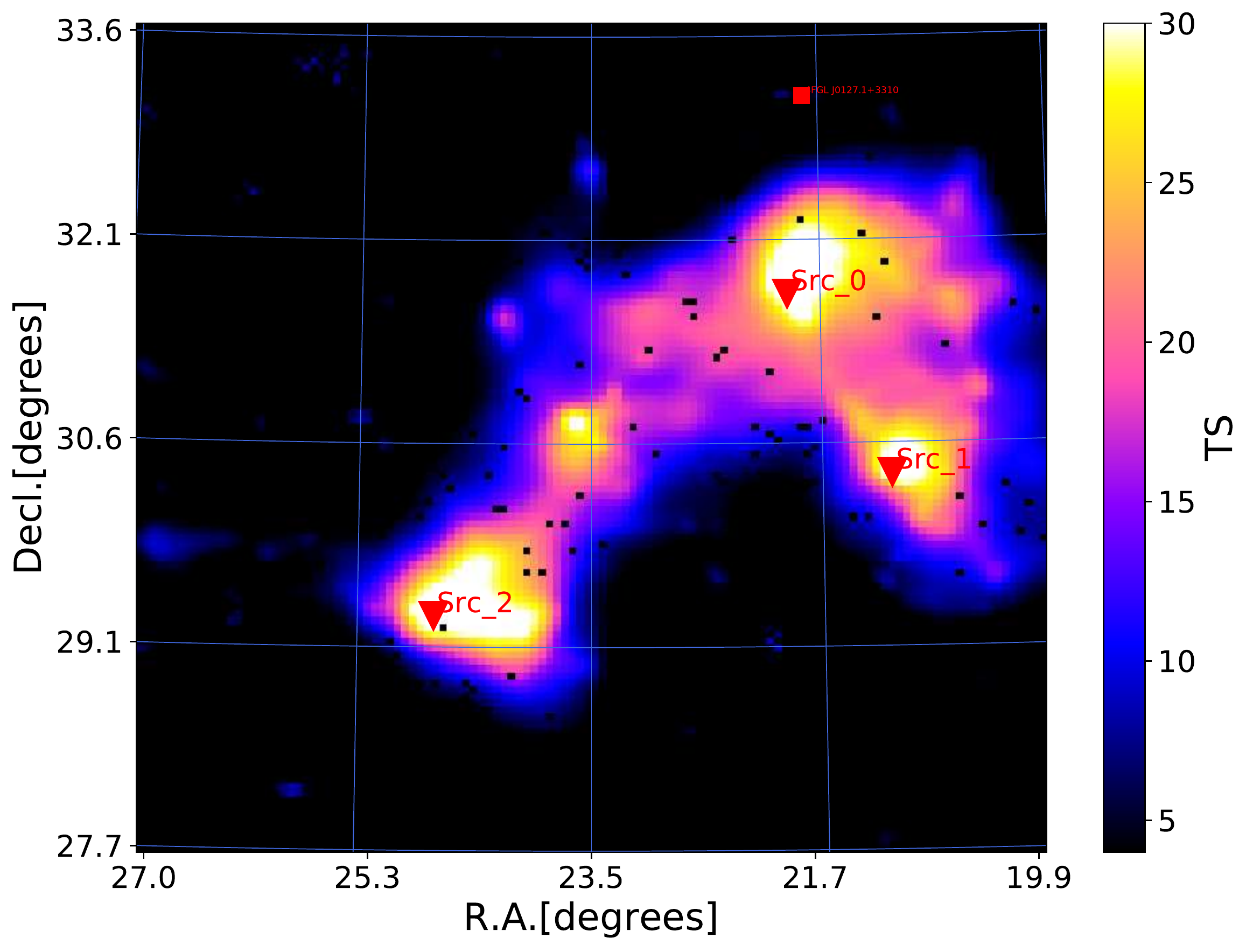}
\includegraphics[width=0.45\textwidth]{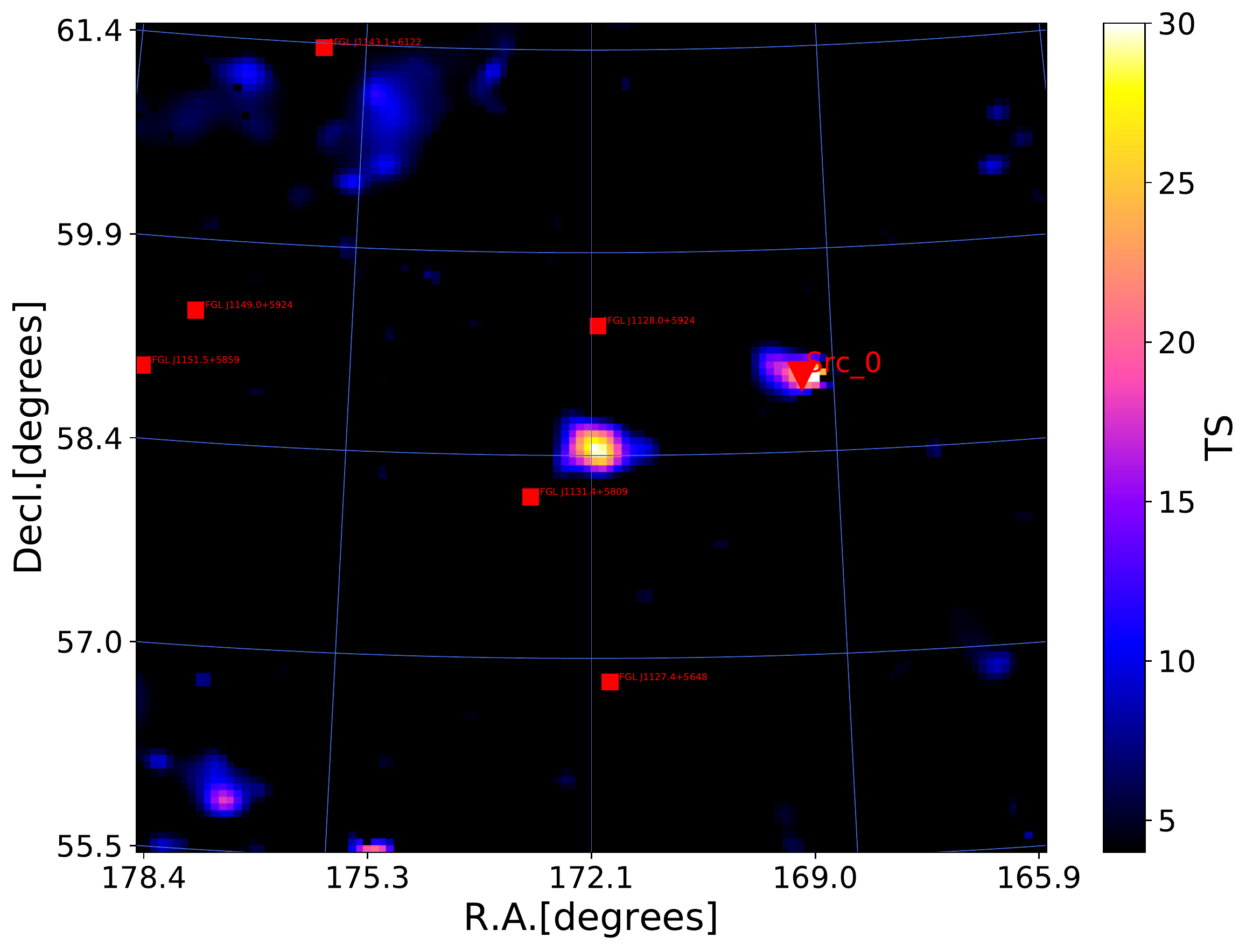}
\caption{$6^\circ \times 6^\circ$ TS map in the energy band $0.3-500$\,GeV for M 33 (left) and for Arp 299 (right). }
\label{fig:8}
\end{figure}

As shown in Fig \ref{fig:8}, we generated a $6^\circ \times 6^\circ$ TS map based on the background modeling combining  the diffuse Galactic emission  and the sources listed in 4FGL catalog.  For the regions around M 33 and Arp 299, we found three and one obvious excesses outside the galaxy radii, respectively. We locate these four new point sources at the positions of the TS  peak and derived their power-law spectral parameter from a broadband spectral fit. The coordinates of four new point sources and their spectral parameters are given in Table \ref{tab:2}.


\begin{table*}[htbp]
\begin{deluxetable}{lccccc}
\tabletypesize{\small}
\tablecaption{\label{tab:2}}
\tablewidth{0pt}
\tablehead{
\colhead{Name} &\colhead{$TS$}  &\colhead{$R.A.$}&  \colhead{Decl.} & \colhead{$N_0$}& \colhead{$\Gamma$}   \\ 
                           &                          &    \colhead{[deg]}&    \colhead{[deg]} &  \colhead{$\times 10^{-14}\rm\ cm^{-2}\ s^{-1}\ MeV^{-1}$}           &    \\
                                         \colhead{(1)} &   \colhead{(2)} &  \colhead{(3)} &  \colhead{(4)} &  \colhead{(5)} &  \colhead{(6)} }

\startdata
\multicolumn{6}{c}{$6^\circ \times 6^\circ$ region around M33}\\
\hline%
Src\_0 & 34.1 & 21.914  & 31.711 & 1.07$\pm$0.39 & 2.75 $\pm$0.28  \\
Src\_1& 29.3   & 21.098 & 30.439 &  0.99$\pm$0.38 & 2.77$\pm$0.26  \\
Src\_2& 38.5   & 24.657 & 29.36 &  1.04$\pm$0.38 & 2.87$\pm$0.23  \\
\hline%
\multicolumn{6}{c}{$6^\circ \times 6^\circ$ region around Arp~299}\\
\hline%
Src\_0& 32.7 & 169.308  & 58.989 & 0.65$\pm$0.21 & 1.71$\pm$0.18 \\
\enddata
\tablecomments{The spectrum of each source is modeled as a power law spectrum $dN/dE=N_0(E/E_0)^{-\Gamma}$, where $E_0$ is fixed to 3 GeV.
(1)Source name;
(2)$\rm TS$ value;
(3) Right ascension J2000
(4) Declination J2000;
(5) Power-law spectral normallization $N_0$;
(6) Power-law spectral photon index;}
\end{deluxetable}
\end{table*}

{To study the impact of different diffuse Galactic emission models, we also created an alternative background model using the old diffuse Galactic emission template \citep[i.e., $\rm gll\_iem\_v06.fits$, ][]{2016ApJS..223...26A}. {   Note that the spectrum of isotropic model is shaped by $\rm iso\_P8R3\_SOURCE\_V2.txt$, which apply to the analysis of Pass 8 events with IRF $\rm P8R3\_SOURCE\_V2$ considering the Galactic model $\rm gll\_iem\_v06.fits$. } We generated the TS maps for M33 and for Arp~299 based on this  background model, which are shown in Fig. \ref{fig:9}. Comparing the new maps  to that shown in Fig.\ref {fig:1}  and Fig.\ref {fig:4}, we find that the the morphology of the $\gamma$-ray excesses from the two galaxies is almost unchanged.

\begin{figure}[htbp]
\includegraphics[width=0.45\textwidth]{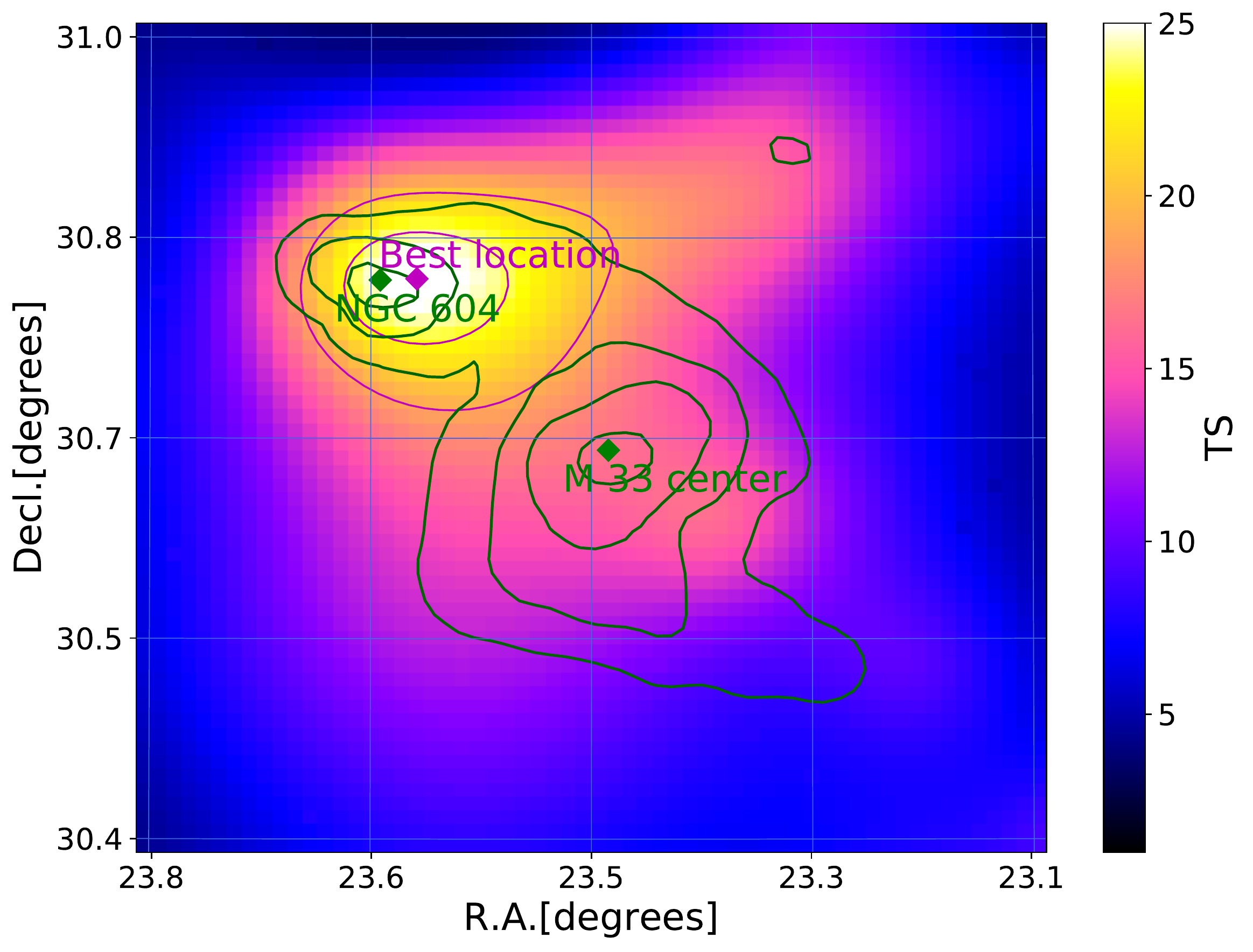}
\includegraphics[width=0.45\textwidth]{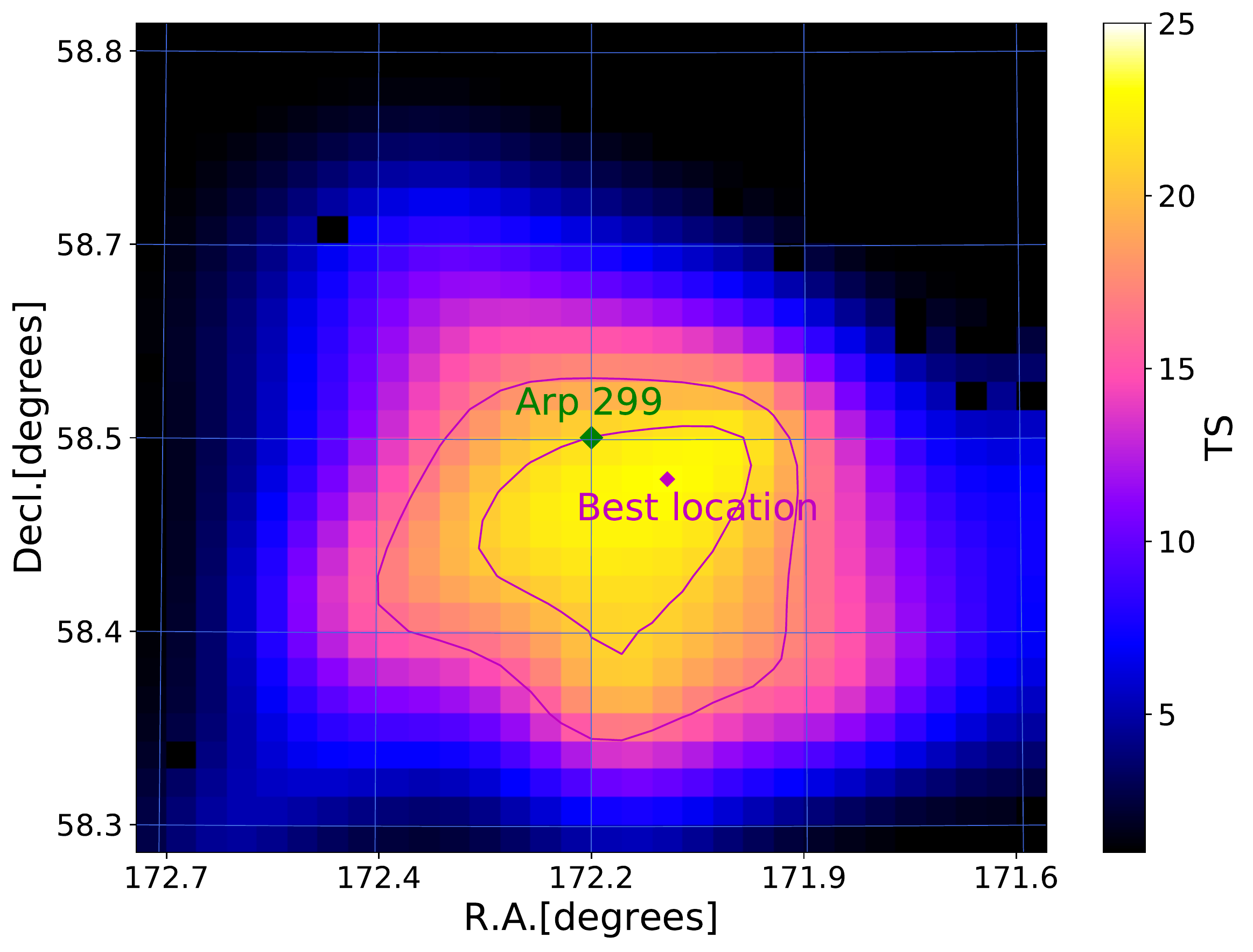}
\caption{TS map in the energy band $0.3-500\,$GeV for M33 (left) and for Arp~299 (right) with background emission subtracted  for the case using the old diffuse Galactic emission template $\rm gll\_iem\_v06.fits$. The dark green contours correspond to measurements of IRAS at $60\mu$m.}
\label{fig:9}
\end{figure}


\end{document}